# HEL1OS – A Hard X-ray Spectrometer on Board Aditya-L1

Anuj Nandi[1], Manju Sudhakar[1], Srikar Paavan Tadepalli[1], Anand Jain[1], Brajpal Singh[1], Reenu Palawat[1], Ravishankar B. T.[1], Bhuwan Joshi[2], Monoj Bug[1], Anurag Tyagi[1], Sumit Kumar[1], Mukund Kumar Thakur[1], Akanksha Baggan[1], Srikanth T.[1], Arjun Dey[1], Veeresha D. R.[1], Abhijit Avinash Adoni[1], Padmanabhan[1], Vivechana M. S.[1], Evangelin Leeja Justin[1], James M. P.[1], Kinshuk Gupta[1], Shalini Maiya P. R.[1], Lakshmi A.[1], Sajjade Faisal Mustafa[1], Vivek R. Subramanian[1], Gayatri Malhotra[1], Shree Niwas Sahu[1], Murugiah S.[1], Medasani Thejasree[1], Narayan Rao G. S.[1], Rethika T.[1], Motamarri Srikanth[1], Ravi A.[1], Nashiket Premlal Parate[1], Nigar Shaji[1]

[1] U. R. Rao Satellite Centre (URSC), Indian Space Research Organisation (ISRO), Department of Space (DOS), Bangalore – 560017, India.
[2] Udaipur Solar Observatory, Physical Research Laboratory, Udaipur – 313001, India.

Email: anuj@ursc.gov.in, ravibt@ursc.gov.in

## Abstract

**HEL1OS** (**H**igh **E**nergy **L**1 **O**rbiting X-ray **S**pectrometer) is one of the remote sensing payloads on board Aditya-L1 mission designed to continuously monitor and measure the time-resolved spectra of solar flares between 8 keV and 150 keV. This broad energy range has been covered by using compound semiconductor detectors: cadmium telluride (CdTe: 8 – 70 keV) and cadmium zinc telluride (CZT: 20 – 150 keV) with geometric areas of 0.5 cm$^2$ and 32 cm$^2$, respectively. A stainless steel collimator provides a field-of-view of 6$^o$ × 6$^o$ optimized to limit the off-axis response while keeping the design within the instrument mass constraints. The in-house designed low-noise digital pulse processing-based front-end electronics has achieved a spectral resolution of ≈ 1 keV at 14 keV (CdTe) and ≈ 7 keV at 60 keV (CZT). The instrument is also equipped with processing and power electronics to process the signal, drive the electronics, bias the detectors with required low and high voltages for optimal performance of the overall system. In this article, we present design aspects of the instrument, results from the pre-launch ground-based tests, and the in-orbit operations, which have indicated optimal performance in line with that expected.

## 1. Introduction

The Aditya-L1 mission (Seetha and Megala 2017) has been developed to conduct synergetic solar observations from the Sun-Earth Lagrangian point 1 (L1), using remote sensing instruments observing in near-infrared (1074.7 nm), optical (530.3 nm and 789.2 nm), UV (200 nm – 400 nm), soft X-rays (1 keV - 30 keV) and hard X-rays (8 keV - 150 keV) and in situ measurements of low (10 eV) and high (5 MeV) energy particles (H, α, heavier ions) and the interplanetary magnetic field (IMF). These instruments will provide complementary measurements to understand the initiation and evolution of coronal mass ejections (CMEs) and their effect in near-Earth interplanetary space in the context of space weather. The remote sensing instruments will also provide insights into the dynamics of the solar upper atmosphere in the context of the drivers of chromospheric and coronal heating. The mission was launched on 2 September 2023, and was inserted into a halo orbit around L1 on January 6, 2024[1]. The halo orbit has a period of 180 days and the expected mission life is around five years.

Figure 1 shows the Aditya-L1 spacecraft (left) with the mounting configuration of seven scientific payloads. The HEL1OS payload is mounted on the intermediate deck (top) towards the +yaw axis of the spacecraft. On the right of the figure, the 3D model of the HEL1OS payload is shown with all its major sub-assemblies marked.

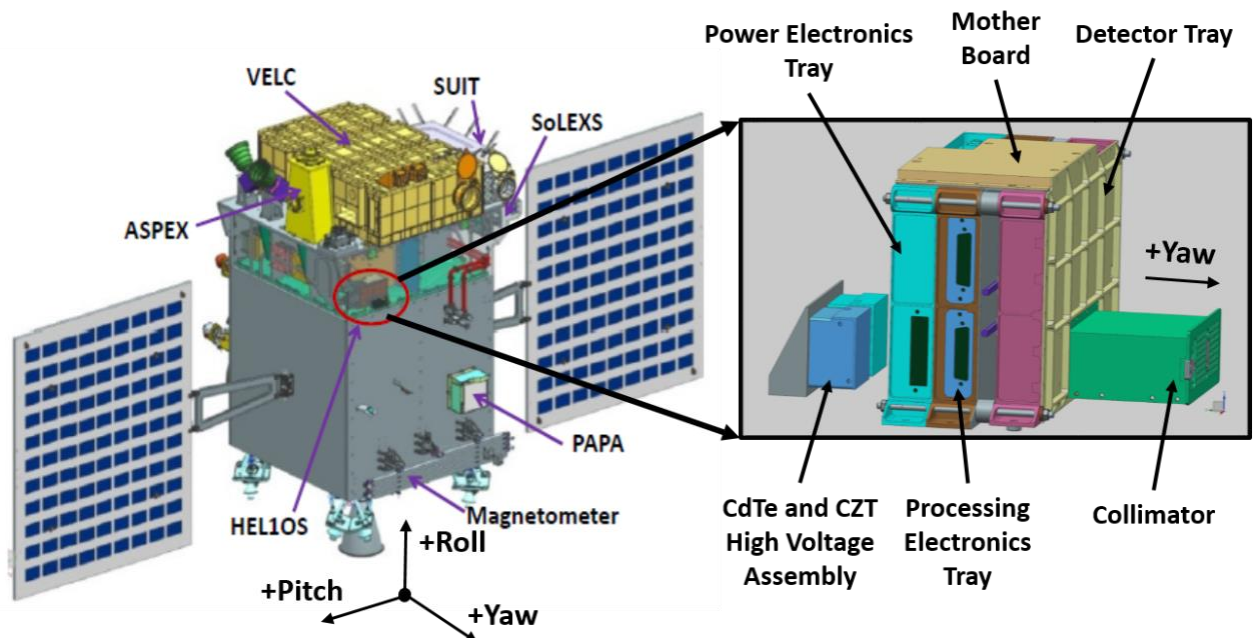

**Figure 1:** The mounting configuration of seven scientific payloads including HEL1OS (on the intermediate deck) on the Aditya-L1 spacecraft (left). Overall 3D model of HEL1OS payload, marked with major sub-assemblies (right). See text for details.

Solar flares are among the most powerful and energetic events in the solar system, releasing vast amounts of energy stored in the Sun's magnetic field. These events accelerate particles, heat plasma, and emit radiation across the electromagnetic spectrum, including intense X-ray emissions. High-energy X-rays (HXR), particularly those above 10 keV, are crucial for understanding the particle acceleration and plasma heating during solar flares. The HEL1OS (right-side of Figure 1) payload, on board Aditya-L1, is poised to advance our understanding of these processes by providing continuous observations in the 8 keV to 150 keV energy range. Operating from the L1 point, HEL1OS offers an uninterrupted view of solar activity, free from the limitations faced by previous missions operating in low-Earth orbit (LEO) or elliptical solar orbits.

One of the most important missions in solar flare research in this millennium, has been Reuven Ramaty High-Energy Solar Spectroscopic Imager (RHESSI) which operated from 2002 to 2018 (Lin et al. 2002). RHESSI revolutionized our understanding of solar flare physics by providing simultaneous imaging and spectroscopic analysis in hard X-rays and gamma-rays, allowing researchers to study thermal and non-thermal processes during flares (Emslie et al. 2011). Among RHESSI's key contributions was the exploration of how non-thermal electrons, which produce HXR emissions through bremsstrahlung, play a dominant role in the energy budget of solar flares (Dennis et al. 2022). However, RHESSI's operation from LEO introduced limitations, such as periodic interruptions due to passage through the Earth's orbital "night-side", interference from the South Atlantic Anomaly (SAA), and latitudinal dependence on the high-energy background.

---

[1] https://www.isro.gov.in/halo-orbit-insertion-adtya-l1.html

Building on RHESSI legacy, the Spectrometer/Telescope for Imaging X-rays (STIX) on board Solar Orbiter (Krucker et al. 2020) is currently delivering high-resolution observations of solar flares, with a focus on flare loop dynamics and the transport of energetic particles. STIX's elliptical orbit enables observations from varying distances and perspectives, offering valuable multi-point data. Using STIX data, Purkhart et al. (2023) provided meaningful advancements in understanding energy dissipation processes and particle transport within flare loops, demonstrating the benefits of multi-instrument and multi-viewpoint analyses in unraveling flare evolution. Studies of flares occurring behind the solar limb (Pesce-Rollins et al. 2024) have uncovered energy transport between primary flare sites and distant on-disk regions, suggesting that energetic particles can travel along large-scale magnetic loops, linking widely separated regions on the solar surface. Volpara et al. (2023) employed a regularized imaging spectroscopy technique with STIX, reconstructing electron flux distributions at multiple energy levels and enabling precise tracking of electron acceleration and transport from the coronal source to the chromosphere. STIX's capability to investigate microflares, originally pioneered by RHESSI (Stoiser et al. 2007; Hannah et al. 2008), remains a critical asset. An investigation of a microflare associated with a coronal jet (Battaglia et al. 2022) pinpointed the energy release site above the flare loop, aligning with the electron acceleration region and supporting interchange reconnection as a mechanism for particle acceleration in the lower corona. Furthermore, STIX observations of microflares in sunspots, where strong magnetic fields enhance electron acceleration (Saqri et al. 2024), continue to refine our understanding of small-scale energy release and particle dynamics.

The Hard X-ray Imager (HXI) on board the Advanced Space-based Solar Observatory (ASO-S) represents the latest advance in high-resolution solar hard X-ray imaging spectroscopy, enabling detailed diagnostics of flare energetics and particle acceleration (Gan et al. 2023). Li et al. (2024) confirmed that 149 HXI flaring events whose impulsive phases were observed adhered to the Neupert effect, reinforcing the link between non-thermal electrons and thermal plasma heating. Shi, F. P. et al. (2024) detected $\approx 27\,\text{s}$ quasi-periodic pulsations during an X1.2-class flare, indicating episodic acceleration processes. Shi, G. L. et al. (2024) also reported strongly asymmetric HXR ribbon emission in a long-duration flare, attributed to loop geometry and magnetic mirroring. Shamsutdinova et al. (2024) combined HXI with radio data to track the co-evolution of microwave and HXR sources during a limb flare, modeling flare energy release and transport. Gou et al. (2024) linked HXI flare signatures to coronal blowout jets, showing sunspot rotation as the trigger for both jet and flare activity. Triangulation studies combining STIX and HXI data (Ryan et al. 2024) have constructed the three-dimensional architecture of flare loops and identified non-thermal electron acceleration regions, offering a nuanced understanding of energy transfer across extended magnetic structures.

HEL1OS's position at the L1 point uniquely complements X-ray imaging observations from STIX by offering a high-resolution spectral observation combined with extended periods of a common observational window, free from Earth shadowing and maintaining a stable background. Unlike RHESSI and STIX, which utilize movable attenuators to manage strong X-ray flux at low energies during intense flares, HEL1OS employs a fixed filter (see Table 3) that ensures time profiles remain consistent despite fluctuations in flare flux. This stability is particularly advantageous for temporal analyses such as quasi-periodic pulsations (QPPs) and in studies requiring robust detrending of light curves. By combining continuous coverage with a design that avoids instrumental-induced variability, HEL1OS enhances our capability to perform precise, high-cadence investigations of dynamic solar flare phenomena.

HEL1OS will be able to detect both high-temperature thermal and non-thermal emissions, providing a comprehensive perspective of solar flare energetics, from acceleration of high-energy electrons to plasma heating. By combining these observations with complementary flare spectra from the Solar Low Energy X-ray Spectrometer (SoLEXS; Sankarasubramanian et al. 2017), also on board Aditya-L1, HEL1OS enables multi-temperature analysis of plasma heating during solar events. A critical parameter derived from HEL1OS's HXR spectra is the low-energy cut-off in the electron spectral distribution, which plays a pivotal role in disentangling the thermal and non-thermal components of the energy release. The synergy between HEL1OS and SoLEXS will refine this constraint by more effectively isolating the contribution of super-hot plasma from non-thermal HXR emissions, enhancing our understanding of energy partitioning during flares.

HEL1OS is also well-poised to study QPPs, a characteristic feature of solar flares observed across HXRs, microwave, and ultraviolet emissions. QPPs are thought to arise from periodic processes such as magnetohydrodynamic waves or episodic magnetic reconnection events (Zimovets et al. 2021; Nakariakov and Melnikov 2009; Shi, F.P. et al. 2024). Previous observations by RHESSI, STIX, and other instruments have demonstrated that QPPs can exhibit multiple periodicities, manifesting during both impulsive and decay phases of solar flares (Inglis et al. 2016; Hayes et al. 2019; Li et al. 2024). For instance, French et al. (2024) reported highly coherent ≈ 50 s pulsations during the impulsive phase of a long-duration M-class flare, observed simultaneously by STIX, Geostationary Operational Environmental Satellite (GOES) X-ray sensor (XRS), and Interface Region Imaging Spectrograph (IRIS). A broader statistical analysis by Szaforz et al. (2025) of 8409 STIX-observed flares identified QPPs ranging from 43 to 1355 seconds in 32% of GOES M-class flares. HEL1OS's uninterrupted data collection capability positions it to explore the complexities of QPPs, contributing to a deeper understanding of the periodic mechanisms driving energy release during solar flares. The continuous coverage provided by HEL1OS will be particularly beneficial for detecting and analyzing fast temporal variations, including those associated with large and spectrally hard flares (Knuth and Glesener, 2020). By capturing the full temporal evolution of QPPs without data gaps, HEL1OS is expected to contribute to the refinement of existing models and to offer valuable data that could inspire further investigations into the underlying physics of these rapid energy bursts, as well as flare-driven particle acceleration and plasma dynamics.

HEL1OS thus builds on the foundations laid by past missions studying solar flares. Its continuous, high-sensitivity observations across the broad energy range of 8 – 150 keV from Sun-Earth L1 will allow it to capture the complete evolution of solar flares without the

HEL1OS on Aditya-L1

interruptions faced by LEO orbiting missions or the varying perspective of highly elliptical orbits. By combining data from HEL1OS with that of SoLEXS on board Aditya-L1 and other contemporary instruments like STIX and HXI, researchers will be able to construct a more comprehensive picture of flaring energetics and dynamics and, thus, will result in an improved understanding of particle acceleration, energy transport and partition between thermal and non-thermal processes.

The manuscript is organized as follows. In Section 2, we present the scientific overview that can be addressed with HEL1OS. The design configuration of the HEL1OS payload is presented in Section 3. In Section 4, we present brief results from ground test and calibration of the instrument. The achieved specifications of the instrument and early in-flight results are presented in Section 5 and 6, respectively. Finally, we conclude in Section 7.

## 2. Scientific Overview

HEL1OS investigation of solar flares mainly focuses on the study of particle acceleration and transport by making use of fast-timing measurements and high-resolution spectroscopy. As part of the Aditya-L1 solar observatory, HEL1OS will provide complementary measurements to other remote sensing instruments like SoLEXS, Solar Ultraviolet Imaging Telescope (SUIT), and Visible Emission Line Coronagraph (VELC) to help provide insights into the trigger processes of solar flares and the early initiation and acceleration mechanics of CMEs. We have divided these broad areas into the following research areas:

1. Evolution of the low-energy cut-off in the HXR emitting electron distribution.
2. QPPs in HXR during the flare impulsive phase to understand its connection with non-thermal electron acceleration and transport mechanisms.
3. Study of pre-flare activity and X-ray precursor phase of solar flares.
4. Flare-CME associations in the context of probing the relation between HXR spectral parameters of flares and early kinematic evolution of associated CMEs.

In the following subsections, the current research on each of the areas listed above is elaborated, and it is also explained how HEL1OS is well suited to contribute to furthering that knowledge.

### 2.1 Low-Energy Cut-Off of the Electron Distribution

The HXR emission observed during solar flares is a consequence of the interaction of energetic (non-thermal) electrons accelerated at or in the vicinity of the magnetic reconnection site with the solar atmosphere via bremsstrahlung. The HXR emitting electron spectrum is typically modeled as a single or broken power law distribution with a low-energy cut-off, $E_C$. The available magnetic energy released during flares sets a practical limit on the number of electrons that can be accelerated and the total energy content contained in these electrons (Aschwanden et al. 2019).

It is critical to establish or constrain the value of $E_C$ because, as discussed in Holman et al. (2011), the energy carried by accelerated electrons is highly sensitive to this parameter. Without a proper estimate of the low-energy cut-off, the total number of accelerated electrons, as well as the energy contained in them could be vastly under or overestimated. This is indicated by the equations,

$$N_{nth} = \int_{E_C}^{+\infty} F_{HXR}(E_{HXR}) \, dE_{HXR} \quad \text{electrons s}^{-1}$$

and

$$P_{nth} = \int_{E_C}^{+\infty} E_{HXR} \cdot F_{HXR}(E_{HXR}) \, dE_{HXR} \quad \text{erg s}^{-1}$$

where, $N_{nth}$ and $P_{nth}$ are the total HXR-producing non-thermal electron rate and power, respectively, $F_{HXR}(E_{HXR})$ is the injected or HXR-producing electron flux distribution, and $E_{HXR}$ corresponds to the energy of the electron that produces HXR emission.

Holman et al. (2011) explained that detecting the low-energy cut-off directly is challenging because thermal X-ray emissions often dominate the spectrum at lower energies, masking the non-thermal component. This makes distinguishing between the thermal and non-thermal contributions difficult, especially in the energy range where the cut-off is likely to occur unless one makes use of advanced spectral analysis techniques and high-resolution data.

Several researchers have attempted to constrain the value of $E_C$ by forward-fitting the observed HXR spectrum using the thick-target model, total electron number model, time-of-flight model, warm target model, and cross-over model (Sui et al. 2005; Kontar 2008; Kontar 2011; Kushawaha et al. 2015; Aschwanden et al. 2019; Sahu et al. 2020; Xia et al. 2021) to vary from as low as ≈ 9 keV to as high as ≈ 50 keV for C, M and X GOES class flares. By combining high-energy spectral data from HEL1OS with low-energy observations from the SoLEXS instrument, it will be possible to better isolate the thermal and non-thermal components of the emission, thereby providing tighter constraints on $E_C$. This synergy will not only improve the accuracy of energy partitioning during flares but also enhance our understanding of the mechanisms driving particle acceleration and plasma heating in the solar corona.

### 2.2 QPPs during the Impulsive Phase of Solar Flares

QPPs are characterized by repetitive bursts of radiation observed across different wavelengths (including HXR, radio, and EUV), and are a widespread phenomenon in solar flares. These pulsations occur during all phases of flares, though they are most prominent during the impulsive phase when particle acceleration is at its peak. The low-energy cut-off could be key to understanding how the energy available evolves for these periodic processes such as QPPs and other oscillatory behavior.

Magnetohydrodynamic (MHD) oscillations (see McLaughlin et al. 2012), particularly in magnetic loops, are among the most widely accepted explanations for QPPs. Different modes of oscillations, including sausage modes, kink modes, and torsional Alfvén waves, modulate the plasma density, magnetic field strength, and consequently the emission from flare-accelerated particles. These oscillations may periodically modulate non-thermal particle acceleration, producing the observed pulsations in HXR emissions.

Another group of models attribute QPPs to quasi-periodic magnetic reconnection regimes, where the reconnection process itself becomes periodic due to external drivers, such as MHD waves or internal instabilities within the current sheet. This can lead to a periodic release of energy, which accelerates electrons in bursts, generating quasi-periodic HXR emissions (Chen and Priest 2006;

Nakariakov et al., 2006; Jelinek and Karlicky 2019). Periodic reconnection often occurs in complex flare environments with interacting magnetic loops or evolving current sheets. These reconnection events are key to releasing stored magnetic energy in solar flares, with the resulting non-thermal particles producing QPPs.

Self-oscillatory processes (see also McLaughlin et al. 2009; Thurgood et al. 2017, 2018, 2019, and references therein) in the flare environment, where a steady energy input (such as continuous magnetic flux inflow) is periodically converted into alternating bursts of particle acceleration, also provide a plausible mechanism for QPPs. This model explains why QPPs often occur in multiple phases of flares and can persist for long durations. Such periodic behavior leads to recurring bursts of non-thermal electron acceleration, manifesting as QPPs in HXR emissions.

Whatever the mechanism or combination of mechanisms, the result is the cyclic acceleration of non-thermal electrons, which results in the pulsed nature of HXR emissions, corresponding to the observed QPPs.

Statistical survey of 181 solar flares (GOES C-class to X-class) by Pugh et al. (2017) reveals that 20% of the flares (that is, in 37 of the total 181 flares) show convincing evidence of QPPs, with detected periods ranging from 10 seconds to 100 seconds and mean period of the detected QPPs was $\approx 20^{+16}_{-9}$ seconds. The high sensitivity and temporal resolution down to 1 second for the largest flares will enable HEL1OS to detect QPPs in the 8 to 150 keV energy range.

### 2.3 Flare Precursor Activity and Non-Thermal Energy Release

The precursor phase of solar flares, also referred to as the "hot onset", is well recognized in X-ray light curves as a distinct yet subtle intensity enhancement from the flaring region before the impulsive phase (Chifor et al. 2007; Joshi et al. 2011; Mitra and Joshi 2019; Hudson et al. 2021). This phase is typically observed in X-ray energies of ≤ 20 keV (see Figure 1 in Joshi et al. 2011), dominated by thermal emission. However, some studies suggest the presence of a minute non-thermal component in X-ray emission during this phase (Chifor et al. 2007; Joshi 2011), providing critical evidence for early-stage particle acceleration. A detailed comparison of the X-ray precursor phase with localized small-scale activities seen in EUV/UV imaging channels suggests that processes occurring during this phase play a crucial role in triggering large-scale magnetic reconnection, eventually leading to an intense and impulsive energy release (Hernandez-Perez et al. 2019; Mitra and Joshi 2019; Sahu et al. 2020). Observations from the high-sensitivity HXR sensors on board HEL1OS are expected to provide deeper insights into the non-thermal processes associated with the precursor phase.

### 2.4 Relationship between CMEs and Non-Thermal Solar Flare Processes

Many studies (see Vrsnak et al. 2004; Temmer et al. 2010; Vrsnak 2016; Joshi et al. 2016) have implied a close connection between the acceleration phase of CMEs and the energy release in solar flares. One of the most significant aspects discussed is the synchronization between the CME acceleration phase and the impulsive phase of the associated flare, which is marked by an increase in hard X-ray (HXR) emissions. There is a direct correlation between the hardness of the HXR spectra and the CME acceleration. Events with harder (higher-energy) HXR spectra tend to correspond to more impulsively accelerated CMEs, suggesting that stronger particle acceleration in the flare is associated with faster CMEs (Vrsnak 2016).

A statistical survey of 37 CMEs flare-related events by Berkebile-Stoiser et al. (2012) revealed strong correlations between the CME peak velocity and non-thermal flare parameters, such as the total number of accelerated HXR-emitting electrons and the total energy contained in them, particularly for energies > 20 keV. Their study demonstrated that the CME peak acceleration is more strongly correlated with the hardness of the electron spectrum, suggesting that CMEs with higher acceleration rates are associated with flares that accelerate electrons to higher energies, resulting in harder spectra. Notably, they found that in 80% of the events analyzed, CME acceleration commenced, on average, approximately 6 minutes before the onset of non-thermal HXR emission.

Table 1: Summary of science design requirements and instrument capabilities.

| Science | Desired Requirements | HEL1OS Capabilities | Remarks |
|---|---|---|---|
| Particle Acceleration | HXR emission > 10 keV with energy resolution ≈ 1 keV (10 – 40 keV). | Energy ranges 8 – 150 keV. Energy resolution ≈ 1 keV at 14 keV and ≈ 7 keV at 60 keV. | Evolution of the low-energy cut-off of the HXR-emitting electron spectrum and its relation with the thermal and non-thermal crossover energy. |
| | Timing accuracy of the order of 10s of seconds to study QPPs (typical pulse duration few minutes). Timing accuracy of the order 1 second for HXR bursts. | Time-tagged event data ≈ 10 msec resolution. Based on flare statistics, binning of the light curves can be optimized. | QPPs and HXR bursts observed in HXR imply particle acceleration. |
| | Angular resolution of few arcseconds. | Not an imaging instrument. | Complemented by SUIT full-disk UV images, GOES, and SDO/AIA full-disk EUV images. |
| Triggering Mechanisms (pre-cursor emission) | HXR observations between 1 and 30 keV with ≈ 1 keV energy resolution. | 8 keV onwards with ≈ 1 keV resolution at 14 keV (8 – 40 keV). | SoLEXS will complement for energies < 8 keV. |
| Relationship between flare and CME evolution | Energy range ≈ 5 keV to 150 keV | 8 keV to 150 keV | Adequate to examine non-thermal parameters and early acceleration of CMEs. |

More recently, Vievering et al. (2023) examined the temporal relationship between HXR bursts and CME acceleration in 12 well-observed events between 2010 and 2013, reinforcing and extending the earlier findings of Berkebile-Stoiser et al. (2012). They reported that HXR bursts predominantly occur throughout the main acceleration phase of CMEs. Additionally, Vievering et al. (2023) found that faster CMEs are more commonly associated with multiple HXR bursts, highlighting a potential link between intermittent magnetic reconnection and CME kinematics.

In conjunction with the observations from UV and coronagraphic imagers on board Aditya-L1, HEL1OS measurements are expected to make important contributions toward understanding the flare-CME associations. In order to achieve the science goals, the instrument requirements are defined for the HEL1OS payload. In Table 1, we present the desired requirements to meet the science objective along with the capabilities of the instrument achieved.

## 3. Configuration Design

HEL1OS instrument is designed and developed to study solar flares in X-rays in the energy range of 8 – 150 keV. In Figure 1 (right side), the 3D model of the HEL1OS payload is shown with all major sub-assemblies. The payload consists of a machined aluminum housing composed of,

- Three mechanical trays:
  front-end electronics (FE) Tray with front-end/detector card and the four detectors.
  processing electronics (PE) Tray consisting of the processing electronics card.
  power tray consisting of power conditioning electronics (PCE) with M3G DC-DC converter and single event latch-up (SEL) mitigation card.
- One sub-assembly (L-bracket) with two high voltage (HV) boxes consisting of HV electronics for biasing the detectors, and
- A collimator fabricated from a monolithic stainless steel block designed for 6º × 6º FOV.
- The FE, PE, and PCE cards are electronically interfaced via a motherboard (MB).

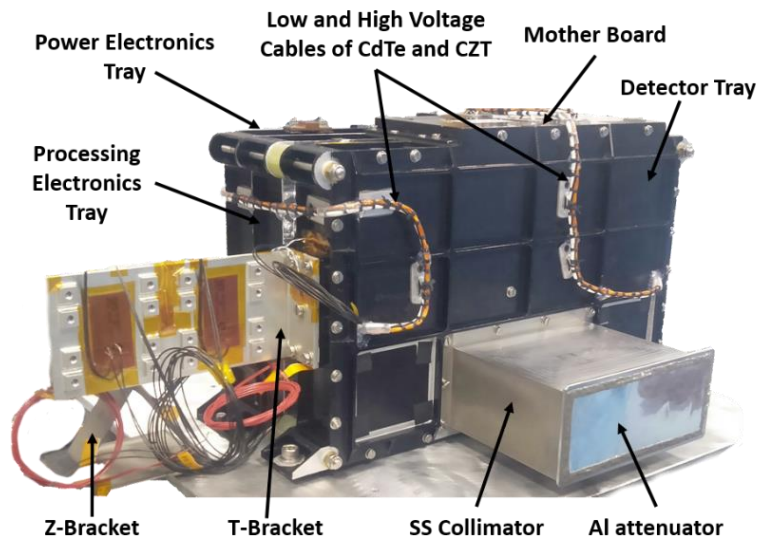

**Figure 2:** Flight model configuration of HEL1OS payload. All major components are also marked.

The desired X-ray energy range is achieved with two different types of semiconductor detectors – cadmium telluride (CdTe, 8 – 70 keV) and cadmium zinc telluride (CZT, 20 – 150 keV). CdTe (Redus et al. 2009) and CZT detectors (Nandi et al. 2009, Kotoch et al. 2011, Bhalerao et al. 2017) are biased with HV of +600 V and -650 V respectively. The output signals from the detectors are processed by the front-end electronics, and subsequently, the processed data are packetized by processing electronics before sending for storage in the on board data handling subsystem. The processing electronics also decodes the relevant telecommand (TC) and sends out the housekeeping (HK) data to telemetry (TM). The Power Electronics is interfaced with the raw bus supply of the spacecraft. The flight model configuration of the payload is shown in Figure 2. Design details of electronics, mechanical, and thermal aspects are presented in Sections 3.2 and 3.3.

It is to be noted that some of the components like the HV generation module from PICO Electronics (Kotoch et al. 2011), the Charge Sensitive Pre-Amplifier (CSPA) from Amptek Inc. and both detectors – CdTe and CZT – are industrial or commercial grade. To ensure space-worthiness of the components and devices, screening and qualification have been carried out at the U. R. Rao Satellite Centre (URSC) of the Indian Space Research Organisation (ISRO) (Jain et al. 2025). In the following sub-sections, we provide some details of various sub-systems of the payload.

### 3.1 Detector System

The detector units are at the heart of the HEL1OS payload. Detectors are solid-state based semiconductor type and are selected to achieve the scientific objective with desired energy ranges, energy resolution and required efficiency.

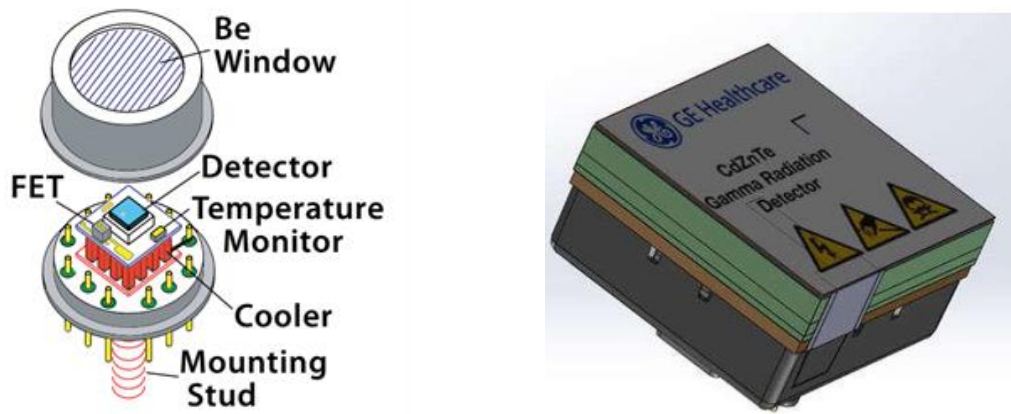

**Figure 3**: Detector devices (CdTe and CZT) used in HEL1OS payload to cover the lower (8 – 70 keV) and higher (20 – 150 keV) energy X-ray ranges, respectively. On the left, an exploded view of the TO-8 package of the CdTe detector is shown, whereas on the right an isometric view of the CZT module with the CZT crystal (shaded in green) along with read-out ASIC (enclosed in grey housing) is shown. CdTe detectors (left) are from Amptek Inc. and CZT detectors (right) are from GE Healthcare.

The CdTe detector is a standard XR-100T unit from Amtek Inc. (Redus et al. 2009) with dimensions $5 \times 5 \times 1$ mm$^3$ housed in a TO-8 package with an in-built two-stage thermoelectric cooler. A mounting stud, which is provided at the bottom of the package, is used to anchor the detector onto a printed circuit board (PCB) and also act as a heat sink interface (see Figure 3). The thermo-electric cooler (TEC) current is controlled to cool the detector and set at the desired operating temperature. The unit is hermetically sealed with a 100 µm beryllium window to ensure light-tightness. The CZT detector (Kotoch et al. 2011) is from GE Healthcare and is a composite module of pixelated CZT crystals (see Figure 3) bonded onto an application specific integrated circuit (ASIC) that generates digital output. The CZT detector active area is shaped as a square grid array of 16 × 16 pixels with a pitch of 2.46 mm. As the intended application of CZT modules is in the field of medical imaging, they are designed to operate at room temperatures (15 – 25 $^0$C). Thus, the CZT detectors do not have an active cooling system built into their modules. A heat sink finger with an M3 thread is provided to passively remove heat generated in the CZT detector during its operation. By construction, the CZT is an open detector and thus is not light tight. To shield CZT detectors from solar optical photons, a light-tight filter made of aluminized Kapton (Prajwal et al. 2019) is draped over the entire collimator assembly, including the mechanical interfaces (Jain et al. 2025).

In Table 2, we present the detailed specifications of both detectors.

Table 2: Specifications of CdTe and CZT detectors.

| *Parameter* | *Specifications* | |
|---|---|---|
| Instrument Type | Spectrometer with solid state semiconductor detectors | |
| Detector Type | CdTe | CZT |
| Number of Detectors | 2 | 2 |
| Detector Crystal Dimensions | 5 mm × 5 mm × 1 mm per detector, non-pixelated detector | 40 mm × 40 mm × 5 mm per detector<br>256 pixels per detector<br>2.46 mm × 2.46 mm pixel pitch |
| Geometric Area | 0.5 cm$^2$ (both detectors) | 32 cm$^2$ (both detectors) |
| Detector Packaging | Hermetically sealed TO-8 package with 100 μm Be window | Open detector without any package |
| Energy Range | 8 keV to 70 keV | 20 keV to 150 keV |
| Energy Resolution | ≈ 1 keV at 14 keV | ≈ 7 keV at 60 keV |
| Bias Voltage | +600 V | -650 V |
| Operating Temperature | -30 $^0$C to -40 $^0$C with built-in TEC | +10 $^0$C to +20 $^0$C with passive cooling |

Table 3: Details of layers of material in the optical path for estimation of the effective area.

| Layers | CdTe | | CZT | |
|---|---|---|---|---|
| | **Material** | **Dimension,** μm | **Material** | **Dimension,** μm |
| Optical Light-tight Thermal Filter | Kapton | 330.2 | Kapton | 330.2 |
| | aluminium | 8 | aluminium | 8 |
| Soft X-ray Attenuator | aluminium | 150 | aluminium | 150 |
| Detector Window | beryllium | 101.6 | - | - |
| Dead Layers | cadmium | 0.2 | polyester | 150 |
| | | | FR4 | 200 |
| Cathode Contact | platinum | 0.2 | indium | 0.5 |
| Crystal Thickness | CdTe | 1000 | CZT | 5000 |

As mentioned earlier the CZT detector is sensitive to optical light, whereas CdTe is housed in a hermetically sealed TO-8 package. To block the optical light entering through the collimator, and to reduce the flux (due to soft X-ray) and background, a light-tight thermal filter (LTF) and aluminum foil (as attenuator) are implemented on top of the collimator (Jain et al. 2025). The effective area of the detector system of HEL1OS payload is estimated based on various layers of materials used, which are summarized in Table 3.

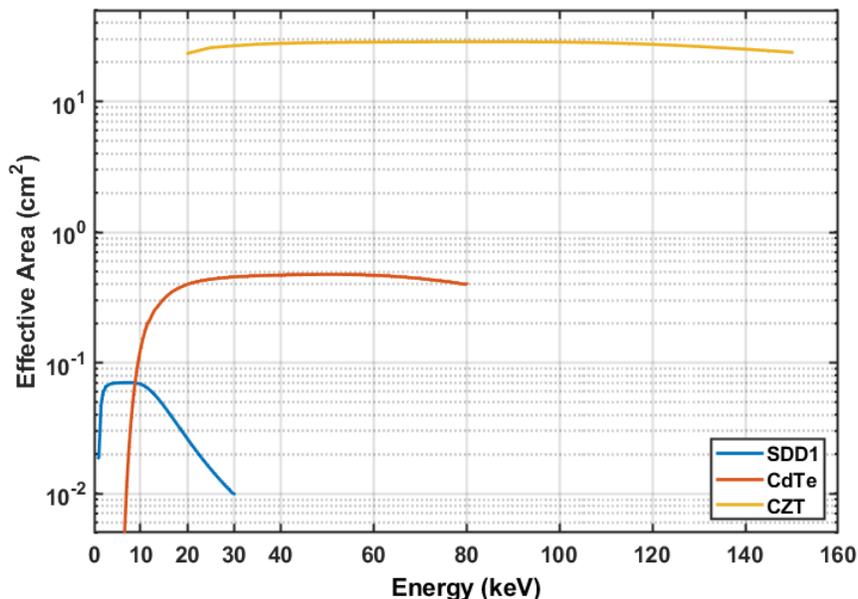

**Figure 4:** Theoretical estimation of the effective area curve of CdTe and CZT detectors (5 keV – 150 keV) plotted along with the effective area of the SDD detector of SoLEXS payload.

The collimator placed in front of the detector assembly reduces the net illuminated area of the detectors. The open area fraction for CZT is around 92%, and that for CdTe is around 98%. The collimator for each non-pixelated CdTe detector has an open area of ≈ 5.1 mm × 5.1 mm to accommodate the CdTe crystal of dimension 5 mm × 5 mm. In comparison, each CZT is a 16 × 16 grid pixelated detector of 256 pixels each, and each pixel is of dimension 2.46 mm × 2.46 mm. The slats of the collimator for the CZT detector have an open area of 4.96 mm × 4.96 mm each catering to 2 × 2 neighbouring pixels, the slat thickness being 200 ± 25 µm. Computation of the effective area is done analytically using X-ray attenuation coefficients from the National Institute of Standards and Technology (NIST) photon cross-section database for the aforementioned materials, thicknesses, and for an illumination angle of $90^0$ from the detector plane. In Figure 4, we show the variation of the effective area of both detectors over the energy range of 5 keV to 150 keV along with the Silicon Drift Detector (SDD) of the SoLEXS payload (Sankarasubramanian et al. 2017). This theoretical estimate shall be corrected after on board observations of standard X-ray sources like the Crab pulsar (Trimble 1968, Coroniti 1990, Hester 2008 and references therein).

### 3.2 Electronics Design

HEL1OS electronics is designed to read out and process the data from both types of detectors (CdTe and CZT detectors), based on the required energy resolution and energy range of operation. The CZT detector module consists of two 128-channel readout ASICs to process the data from 256 pixels of the detector. The ASIC consists of CSPA, shaper, and analog to digital converter (ADC). The digitized data is stored in a local buffer along with pixel information. The module also has configuration registers to set the threshold, shaping time, pixel enable/disable, etc. The detector module provides a serial peripheral interface (SPI) to read out the data as well as to configure the parameters like energy threshold, enabling/disabling of pixels, etc. (Kotoch et al. 2011, Sreekumar et al. 2011). The highly sensitive analog signal from the CdTe detectors is processed using a CSPA followed by post-amplification, anti-aliasing filtering, and digitization. The Digital Pulse Processing (DPP) is used to further process the data and extract signal information with a desired noise performance (Knoll 2000). The DPP gives a digital value for every X-ray event proportional to the X-ray photon energy. The electronics also provides a selection logic to filter out invalid events. Additional functionalities of the electronics include: maintaining detector temperatures within the operational range, control of high voltage generation and, providing safety features to auto-shut off the vulnerable sub-systems in case of contingencies, details of which are discussed subsequently. Telecommand, telemetry and, data interface to the satellite main frame are also handled by the processing electronics. Figure 5 shows the block diagram of the overall electronics design of the HEL1OS payload.

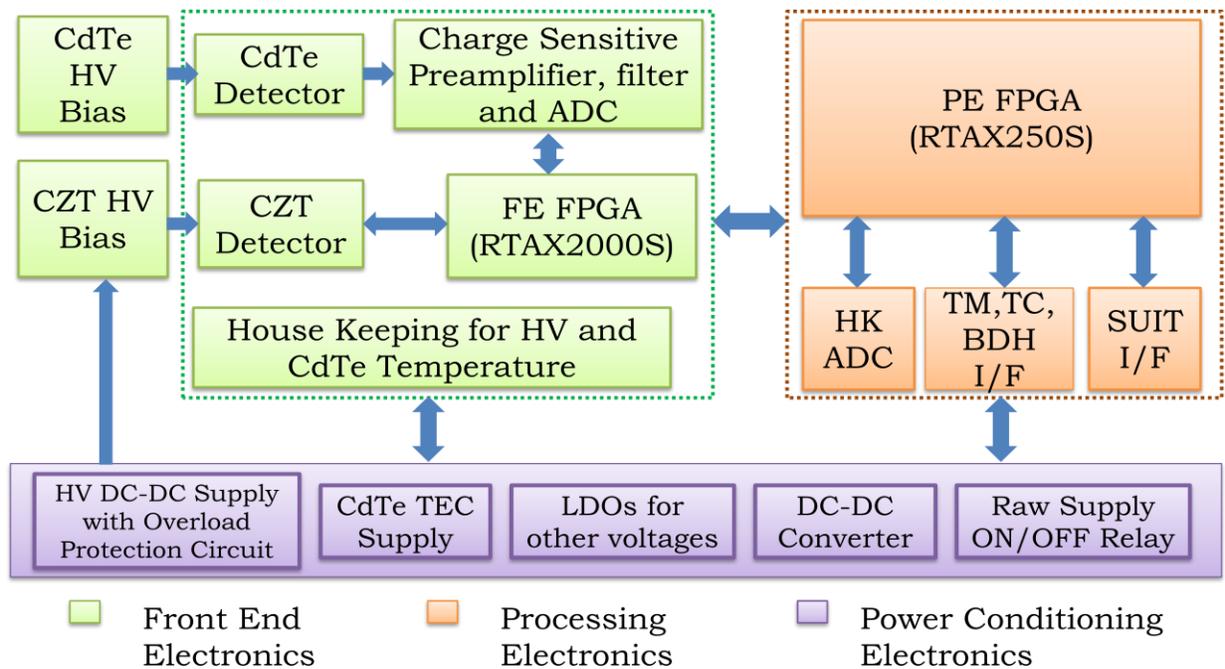

**Figure 5:** Overall electronics block diagram of HEL1OS payload, showing all major functional blocks, including interfaces with other units such as telemetry (TM), telecommand (TC), base data handling (BDH), and SUIT payload. The power conditioning electronics includes the high voltage (HV) supply, supply to CdTe thermoelectric cooler (TEC), low drop-out (LDO) voltage regulators, DC-DC converter and on/off relay.

The overall electronics (see Figure 5) consists of the following hardware blocks: (i) front-end electronics PCB (FE, green shaded block), (ii) processing electronics PCB (PE, orange shaded block), (iii) Power Conditioning Electronics PCB (PCE, purple shaded block), and (iv) HV bias generation PCBs for CdTe and CZT detectors. FE, PE, and PCE are connected through a motherboard (MB) as shown in Figure 2. Outputs of HV bias generation boards are interfaced with FE directly through dedicated HV wires. Some of the electronic design aspects of various blocks are presented below.

- **Front-end electronics PCB**: It comprises the detectors (CdTe and CZT), pre-amplifiers, ADC driver, operational amplifiers (Op-amps), ADCs, and DPP in a Field Programmable Gate Array (FPGA, RTAX2000S). The block-level details of the FE design are shown in Figure 6. The CdTe detector modules have the CdTe diode and field effect transistor (FET) whose output is fed to an external CSPA. The charge integrating feedback capacitor which defines the charge to voltage gain and the feedback resistor to discharge the feedback capacitor are also present inside the CdTe module. The CdTe diode and FET are cooled using a TEC. The TEC voltage is adjustable to achieve the desired temperature. The CSPA output is amplified, filtered, digitized, and sent to the FPGA for DPP. The DPP modules consist of slow and fast triangular shapers, pulse selection logics for pile-up and saturated pulse rejection, and peak detection of valid pulses. The fast shaper has a peaking time of 500 ns and is used for timing trigger and the fast channel peaking time is adjustable in the set [1 µs, 2 µs, 4 µs]. For the CZT modules (Nandi et al. 2009, Kotoch et al. 2011, Sreekumar et al. 2011), the FPGA is configured as the master in the SPI interface and CZT modules as a slave. The SPI interface clock is set to 10 MHz and SPI signals travel through 100 Ω impedance lines as low voltage differential signaling (LVDS) signals. The communication to CZT is initiated by FPGA by issuing a command followed by write, read, or event read cycles. Write and read cycles are used to set or read a configuration parameter such as the low energy threshold, temperature, etc. The event read cycle is used to read the X-ray event registered in the CZT module. Other than these, the FE also consists of HK Op-amps, interfaces with PE to send the detector and HK data, and receives the telecommands for setting the configuration parameters for other operations.

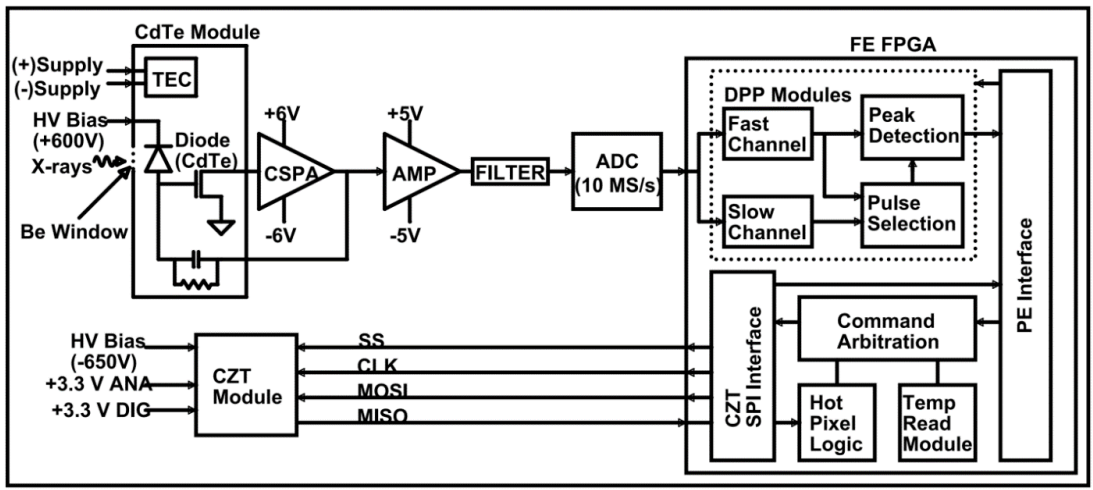

**Figure 6:** Overall front-end electronics (FE) block diagram of HEL1OS payload. All major components are shown.

- **Processing electronics PCB**: PE receives the detector and HK data from the FE. The data is formatted and sent for on board storage. The HK data received from FE along with the HK data generated in the PE are sent for telemetry. The PE also receives the tele-commands from the spacecraft mainframe, which are bifurcated and routed to the FE. The CdTe count rate at per second cadence is sent to the SUIT payload (Tripathi et al. 2023) to be used for on board flare detection logic. All these tasks are performed by the FPGA (RTAX250S) in the PE as indicated in Figure 5.

HEL1OS on Aditya-L1

- **Power conditioning electronics PCB**: It consists of a DC-DC (M3G) converter, relay, low drop-out (LDO) voltage regulators, CdTe-TEC voltage controller circuit, and corona logic circuit to disable HV in case of arcing (more details are explained as part of design features below).
- **HV bias generation PCBs**: They consist of an HV DC-DC converter module, followed by Zener regulators and R-C filters. HV cables are used to connect the HV bias output to the detectors in the FE electronics as shown in Figure 5. The HV bias generation circuit can be controlled by switching the HV DC-DC converter module supply.

Apart from the major functionalities of the hardware blocks, which are mentioned above, some of the important and intricate design features of the electronics along with challenges are mentioned below.

- The front-end electronics PCB is a mixed signal PCB with different types and levels of signals and voltages. High voltage bias (+600 V and -650 V) for biasing the two detectors needs special design care to maintain minimum trace distances to avoid dielectric breakdown and corona discharge. The CSPA output is a maximum of 10 mV for 60 keV X-ray photons. For SPI interface at 10 MHz with the CZT detector module, LVDS lines are used for making FE a controlled impedance PCB.
- FPGA has an SPI interface with CZT detectors to configure the detector parameters and to read the detector data and temperature values using commands (Kotoch et al. 2011, Sreekumar et al. 2011). Some of the commands are received from the ground via telecommand, while some other commands are generated on board by the FPGA modules like the temperature read module and hot pixel disable module. A Command Arbitration Module for CZT has been implemented inside the FPGA to handle the commands generated from these different sources at random times.
- Power ON sequencing is implemented to switch on high voltage with a 500 ms delay after low voltages are on. The current monitoring circuit is implemented on the HV generation PCB to detect any corona discharge or any other malfunction in the HV area of the PCB. HV will be auto shut off in case the corona monitoring circuit triggers and following such an event, it needs to be switched on again by telecommand. The current limit for the corona trigger is set to around 1.8 times the normal current, and the corona trigger status is sent via telemetry. The corona trigger circuit is disabled for some time during the HV on stage to allow for the initial surge current taken by the HV bias generation circuit. A circuit is designed for monitoring of high voltage. High voltages will auto shut off in case the high voltage violates the set limits. The HV will also auto shut off if the temperature of the detectors crosses the set limits.
- On board data processing and intelligence are implemented to reduce the payload data volume. A bunched pixel rejection logic implemented in PE FPGA will reject the CZT events recorded within a time window of less than 6 μs. This logic helps to reject the charge particle events that are recorded within this time window, but (a) spread across multiple nearby pixels or far apart within a CZT module, or, (b) recorded simultaneously in the two CZT modules. The logic looks for coincidence (or bunching) of events in time and not spatial bunching of hit pixels. It is aimed to filter the cosmic ray generated events to suppress the background. These events are found to occur not necessarily between neighboring pixels. The rejection window of 6 μs is selected based on the expected count rate and maximum event readout rate. The saturated event rejection logic rejects the events with maximum ADC value. The hot pixel disable logic is implemented in FE FPGA to remove false/spurious/noisy events from the CZT detector module. If a particular pixel is triggered very often then that pixel is treated as hot and disabled. Piled-up events in CdTe detectors are rejected by the digital pulse processing module in both timing and spectral channels. The details of the ground tests done irradiating CdTe detectors with an X-ray source at different values of incident count rates, the fitting of dead time models to this experimental data to correct for the effective area, are described elsewhere (Srikar et al., in preparation).

- The CZT detector is not a rad-hard device and hence it is prone to radiation effects. A Single Event Latch-up (SEL) mitigation circuit is implemented which senses the increase in the current drawn by the CZT detector in case of latch-up and power recycles the CZT. The current limit for this logic is set to 80 mA, while the nominal current is 40 mA. This avoids the excessive heating and malfunction of the device. The HV DC-DC converter is another non-rad-hard device, which is known to be vulnerable in the case of Single Event Transient (SET), when the HV output may go beyond the safe operating limits. The HV output is therefore monitored and HV is auto shut off in case it violates the set limits.

### 3.3 Mechanical and Thermal Design of the Detector System

The overall instrument design of HEL1OS is centered on two different types of detectors as mentioned in Section 3.1. The overall mechanical design configuration is optimized and validated with detailed Finite Element Analysis (FEA). Overall mechanical integrity of the system is simulated to withstand the launch vibration qualification and stiffness requirements.

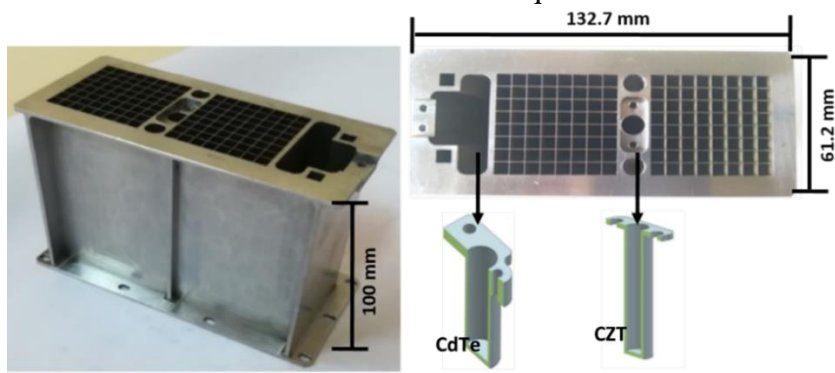

**Figure 7:** The collimator fabricated from a single monolithic SS block (left), showing the mesh of the collimator (right-top) along with a sectional view of radioactive source holders (right-bottom: CdTe and CZT).

The HEL1OS mechanical design and developed flight model are shown in Figure 1 and Figure 2, respectively. A collimator with a FOV of $6^0$ ($\pm\ 3^0$) and a height of $\approx$ 100 mm is designed considering the stringent alignment requirements and ease of fabrication feasibility. The fabrication of the collimator is done with a single monolithic stainless steel (SS) material. Figure 7 shows the fabricated collimator whose open area was designed considering the geometric area of the detectors and their placement on the card. The X-ray radioactive source holders were also fabricated for on board calibration of the detectors using a radioactive source (Am-241), and are an integral part of the collimator.

For optimal performance of the detectors, a dedicated Thermal Control System (TCS) is designed as per the requirements mentioned in Table 2. Figure 8 highlights various thermal components and a specially design thermal-nut for interfacing the CdTe detector. The two sets of detectors (CdTe and CZT) are directly mounted on the detector tray. With an overall ≈ 6 W of heat dissipation from the detector tray which includes dissipation of ≈ 0.7 W and ≈ 0.5 W from CZT and CdTe detectors respectively, a special heat dissipation mechanism was evolved and implemented. As shown in Figure 8, two single-core mini-heat pipes (surface mounted on the tray) were used as a heat transport element from the detector tray to the dual-core heat pipes via a thermal T-clamp, which finally dissipates the heat to the external radiator plate integrated with the spacecraft. The CZT detector base has a cubical mounting (cold finger) interface with a threaded hole. This mount is fastened directly to the tray using a screw. The CdTe detector has a stud to mount and remove the heat. A special mounting and heat extraction device called thermal-nut, made of copper, facilitates heat transmission from the CdTe detector to the detector tray. Provision of lock bolts is made to lock the thermal-nut to the detector tray. For CdTe detectors, in addition to the thermal design system, the internal temperature of the detectors is maintained by varying the voltage of the in-built TEC as mentioned in Section 3.2.

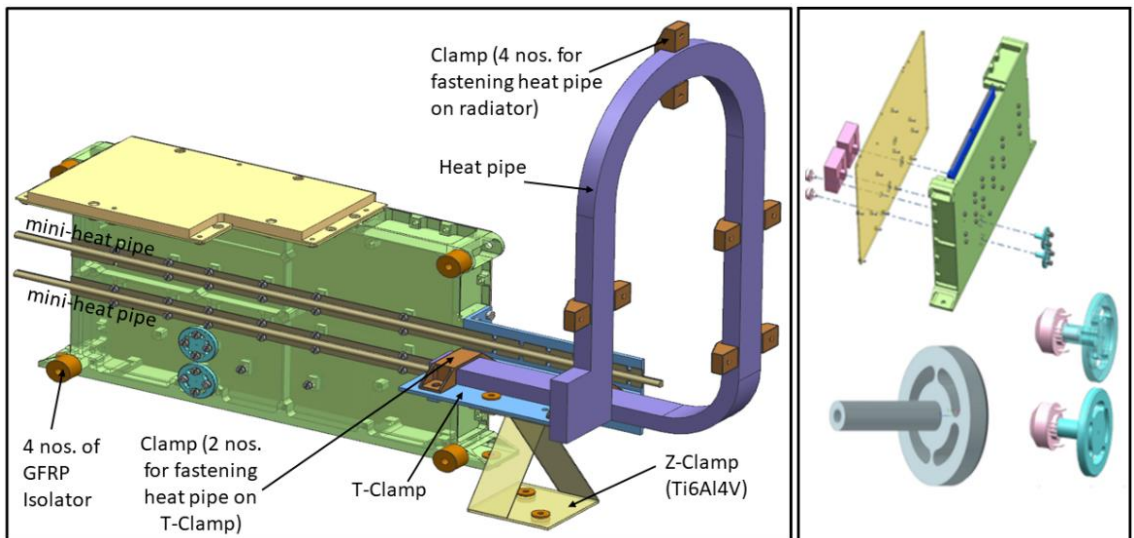

**Figure 8:** Integrated thermal components on the back side of the detector tray (left). Assembly aspects of the CdTe detector with specially designed thermal-nut along with a single unit of thermal-nut for the CdTe detector (right).

## 4. Ground Test and Calibration

The payload has been extensively tested on the ground to determine the spectral response, for different space environmental conditions and to obtain calibration parameters as a function of temperature. Firstly, each of the subsystems was individually tested to ensure optimal performance. The radioactive source (Am-241 with emission lines at energies 13.9 keV, 26.3 keV and, 59.5 keV) is used as the de facto stimulus in all the functional tests on the ground. It was supplemented with other gamma sources like Ba-133 (81 keV), and Co-57 (122 keV, 136 keV) to obtain more emission lines in the higher energies in the CZT detector energy range. Irradiating each detector with this source yields a spectrum in the space of the electronic channels. Associating observed lines in the spectrum to their energies gives a calibration relation for that detector at the given operating conditions. The width of the spectral lines is estimated by

Nandi et al.

curve-fitting to derive energy resolution in terms of full width at half-maximum (FWHM). In Figures 9 and 10, we plot the CdTe and CZT detector spectra irradiated with Am-241 radioactive source. Identified line energies are marked and shown in the channel space of the spectra for both detectors.

After establishing that the spectral output parameters from individual detectors are satisfactory, the payload was integrated and tested under various space environmental conditions. These tests included the initial bench test (IBT), vibration test, thermo-vacuum test, thermal balance test, optical light-tightness test, electromagnetic interference tests, and the final bench test (FBT) before delivery of the payload for integration with the spacecraft. During each of these stages, the functionality of the payload was tested and the key parameters such as the spectral resolution, background count rate, gain, and offsets were monitored.

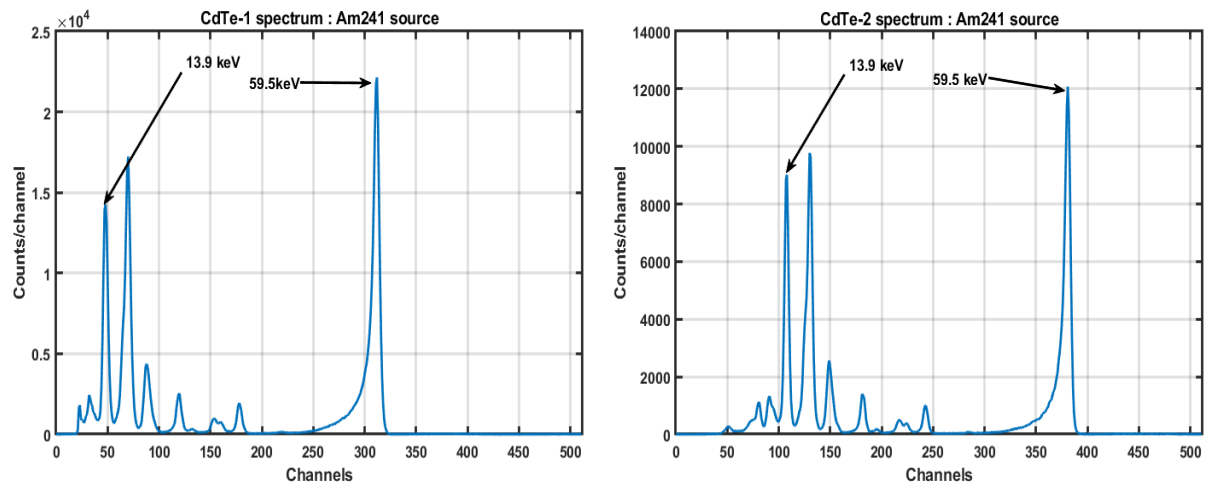

**Figure 9:** Spectra of CdTe-1 and CdTe-2 detectors irradiated with Am-241 source during ground tests. Line energies at 13.9 keV and 59.5 keV are also marked.

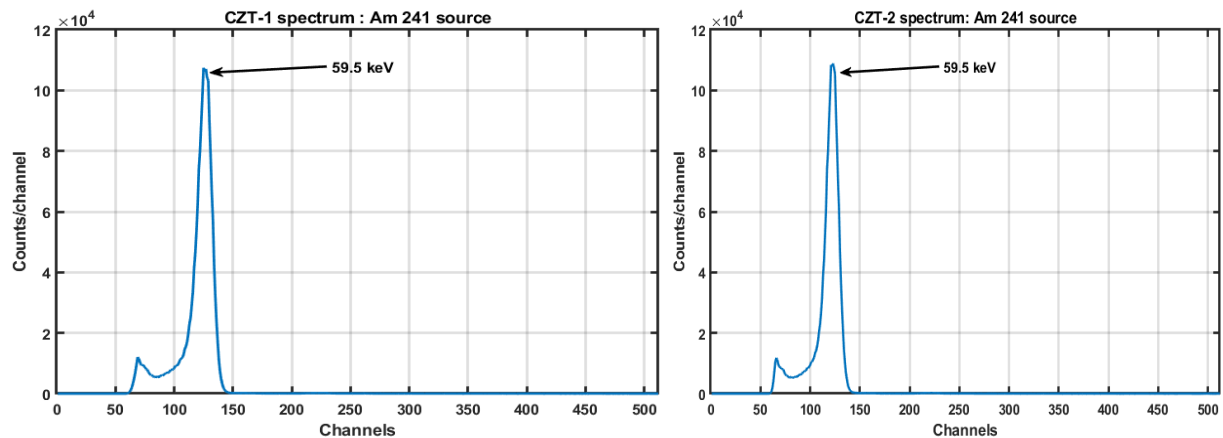

**Figure 10:** Spectra of CZT-1 and CZT-2 detectors irradiated with Am-241 source during the ground tests. Line energy at 59.5 keV is also marked.

In addition to the aforementioned integrated payload tests, the instrument was held at fixed temperatures spanning the on board operating temperature range to irradiate the detectors with multiple radioactive sources with emission lines within the operating energy range. The experimental output is used to obtain the calibration parameters as a function of temperature. Experiments with an active X-ray source are also done to estimate the calibration parameters as a function of energy. At high count rates (> 50 kcps), due to pileup effects, the observed spectrum undergoes appreciable distortion in its shape when more than 5% of events at higher channels are contaminated by pileup events. To ensure the validity of the observed count rates from CdTe detector electronics, tests were conducted to irradiate a standard CdTe (with electronics) module and the CdTe detectors housed in the instrument simultaneously from a quasi-parallel X-ray source. The count rate from the standard CdTe detector (which was initially irradiated with a calibrated Am-241 source of known line flux for the 59.5 keV line) is considered as a reference to compare the obtained count rate of the CdTe in the payload. This test established the efficiency of electronics considering that both detectors – the flight CdTe and the standard CdTe – are identical. As the CZT detector contains within itself the entire signal processing electronics, a similar test with a reference detector is not required and hence not carried out. In Figure 11, we show the device-integrated spectra of CdTe and CZT detectors at various temperatures. The spectra are obtained by setting the instrument base temperature at discrete points such that the device temperatures stabilize to proportionate values. The CdTe detector temperature is varied using in-built TEC, whereas the temperature of the CZT detector is controlled passively. CdTe and CZT detectors are irradiated with Am-241 (13.9 keV, 26.3 keV, 59.5 keV), Co-57 (122 keV, 136 keV), and Ba-133 (81 keV), respectively. Figure 11 shows that there is temperature dependence for CdTe detectors, but the same is not true for CZT detectors. The range of detector temperatures shown in the plot is indicative of the temperature that we expect on board. The current on board CdTe-1 temperature is ≈ -40 $^0$C, while CZT-2 is at ≈ 22 $^0$C. The shift of the CdTe spectrum with temperature is expected and has been calibrated. The obtained calibration relation is incorporated in the data pipeline for converting the channel to energy for detected events.

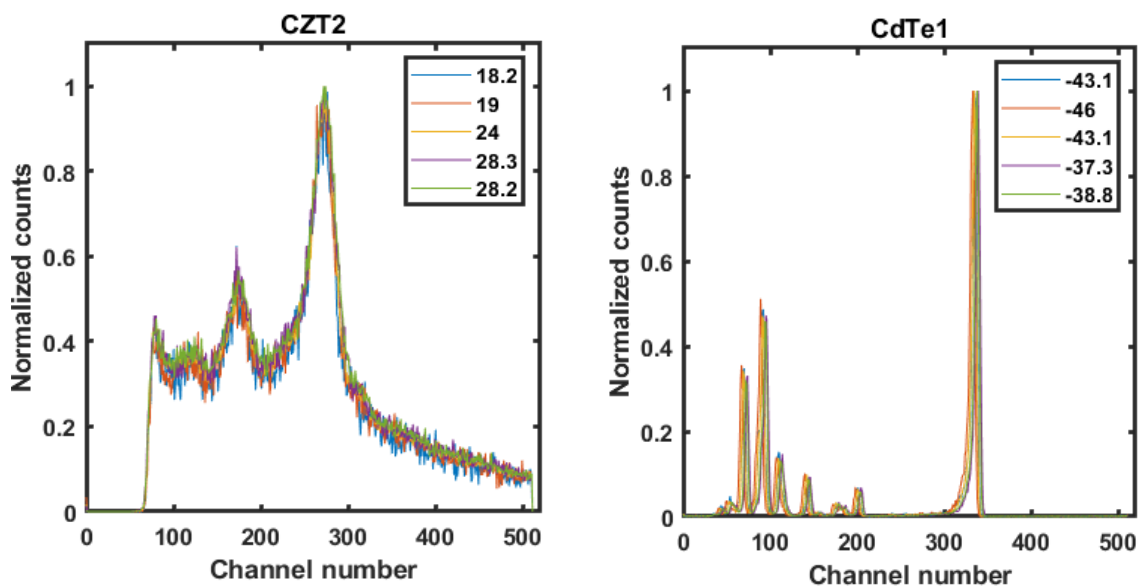

**Figure 11**: Device-integrated spectra for each of the CdTe and CZT detector pairs for different temperatures (in $^0$C as indicated) within the operating range. The CdTe detector is irradiated with Am-241 source and the CZT detector is irradiated with Co-57 and Ba-133 sources.

The spectral response for CdTe detectors has been experimentally obtained, by testing the integrated instrument at INDUS-2 synchrotron beamline at the Raja Ramanna Center for Advanced Technology (RRCAT) facility, in the energy range of 9 – 27 keV. Spectra from the beamline are combined with spectra of other laboratory experiments to generate the spectral response matrix for the instrument. The response for the CZT detector is based on the model developed for the CZTI instrument on board AstroSat (Chattopadhyay et al. 2016). Details on the calibration aspects of the HEL1OS payload will be presented elsewhere (Srikar et al. in preparation). The calibration paper details the aspects of ground testing, simulations performed to estimate the background and collimator response, the detector response generation method, on board background, quantification of on board performance of the system, and verification with contemporary solar X-ray spectrometers.

## 5. Achieved Specifications of the Instrument

HEL1OS has been subjected to extensive calibration on the ground. The initial observed parameters during in-flight operation of the performance verification (PV) phase after launch are quite similar to the ground calibration, details of which will be presented elsewhere (Srikar et al. in preparation). The summary of the achieved specification of the payload is presented in Table 4.

Table 4 – Achieved specification of the HEL1OS instrument.

| Parameter | Specifications |
|---|---|
| Mass (kg) | Payload – 6.1, TCS - 2.0<br>Total – 8.1 |
| Dimension (each in mm) | $232.5 \times 335 \times 194.5$ (Length $\times$ Breadth $\times$ Height)<br>Length: Sun pointing (+ yaw axis) |
| Power (W) | 14.55 |
| Energy Range (keV) | 8 – 150 (CdTe: 8 – 70 keV; CZT: 20 – 150 keV) |
| Energy Resolution (keV) | $\approx$ 1 keV at 14 keV, $\approx$ 7 keV at 60 keV |
| Time Cadence | 10 ms |
| FOV | $6^0 \times 6^0$ |
| FOV Clearance | $\pm 30^0$ |
| Collimator Type | Mesh type |
| Pointing Accuracy | Better than $\approx 1^0$ ($\pm 0.5^0$, provided the reference axis is pointed towards the Sun centre) |
| Operating Temperature | + 15 $^0$C to + 25 $^0$C (CZT)<br>-40 $^0$C to -30 $^0$C (CdTe)<br>+ 10 $^0$C to + 15 $^0$C (Detector Tray) |
| Special Mounting | Thermal isolation for Detector tray to maintain the desired temperature for Detectors |
| Peak Data Rate | 4 Mbits / second |
| Data Volume (max.) | 500 MBytes / day |

## 6. Early In-flight Operation of HEL1OS: First Results

HEL1OS was switched on in the fourth week of October 2023. Detector units on HEL1OS have ever since been capturing the hard X-ray flux of the Sun. The temperature values of the detector tray and the electronics base stabilized to the range of values of ≈ 11 – 13 $^0$C and 21 – 23 $^0$C, respectively, within an hour of switching on the instrument. In comparison, the temperature values of the two detectors reached acceptable operational values within a few minutes of switch on. While the temperature of the CdTe-1 detector has been at about -40 $^0$C, that of the CdTe-2 detector has been at about -28 $^0$C ever since. Likewise, the temperature of the CZT-1 detector has been in the range of ≈ 17 – 18 $^0$C, and that of CZT-2 has been at ≈ 22 $^0$C. During the first few months after the switching on, the data was studied carefully in order to arrive at the final operating configuration gradually for the two detectors. Eventually, the different tunable parameters of the two detectors, energy threshold values, and all essential on board logic were set by the end of June 2024. The CZT detector energy threshold is reduced from its default value of 40 keV, in steps, to 20 keV. The pileup width threshold is adjusted to an optimal value such that noise events with widths greater than the set threshold are filtered out. The CdTe detector timing channel energy threshold is reduced such that it is aligned with the spectral channel energy threshold. A validation measure for the number of recorded events is that at low source flux, the number of events in the spectral and timing channel must be the same within error. Details of on board calibration aspects of the instrument will be presented elsewhere (Srikar et al., in preparation).

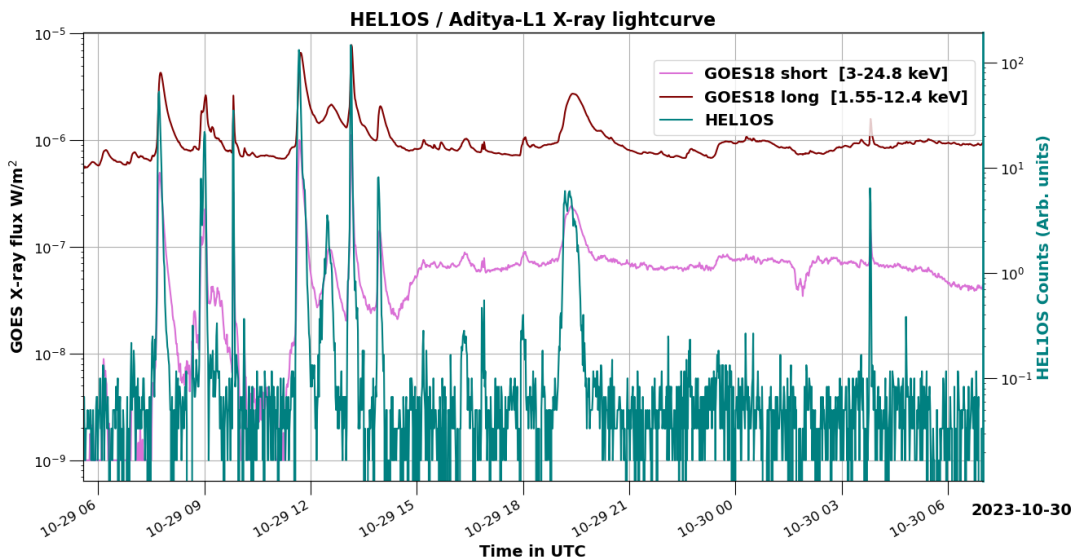

**Figure 12**: First light curve (in green) obtained after switch on of HEL1OS payload on 29 October 2023. Also plotted, with a different y-axis, are the long and short wavelength band light curves from the GOES/XRS instrument for the same duration.

Once the instrument settings are tuned, payload health parameters like temperature and high voltage are continuously monitored for their stability and compared against the ground observations. Radioactive sources (Am-241) mounted within the payload are used to detect and obtain changes (if any) in spectra filtered by non-flaring times. These quiet-time spectra are useful in extracting on board spectral resolution variations that are caused due to factors that are both internal and external to the payload. A database of the derived spectral parameters for

Nandi et al.

various on board conditions is to be generated for updating or selecting an appropriate set of calibration parameters (gain, offset, and energy resolution) for science analysis of a given flare detected by the instrument.

Initial observations of detector-wise light curves and spectra indicate that the net on board performance of the instrument is like its expected performance. The spectral resolution, in terms of FWHM, achieved on board is ≈ 1.2 keV and 7 keV for CdTe and CZT detectors respectively, at 59.54 keV (Am-241). The CdTe detectors are configured to have an energy range of 8 – 70 keV, while the CZT detectors are set for 20 – 150 keV. Light curves during non-flaring times indicate the extent of background in the detectors of the instrument, but a non-solar background can be obtained by steering the spacecraft away from the Sun. This is done during several off-Sun pointing operations during the PV phase of the mission. By collecting the data from all those durations when the Sun is sufficiently away (> 10º) from the boresight, background light curves and spectra are generated. Background count rates of ≈ 0.15 cps and ≈ 70 cps are observed in the light curves of CdTe and CZT detectors respectively.

After switching on, there have been several solar flares that are observed during the cruise phase and post-L1-orbit insertion. Light curves containing flares are compared with GOES broadband flux light curves to bring out the sensitivity of the detectors. Absolute comparison in source flux is not feasible as the instruments operate in different energy bands, and cannot generate data products in a common energy band. In Figure 12, we provide a comparison of observed uncorrected count rates for different GOES classes of flares. Counts enhancement in the CdTe detector of HEL1OS is seen to match with the C-class flare profiles observed with GOES. It indicates that CdTe detectors of HEL1OS can detect down to B-class flares.

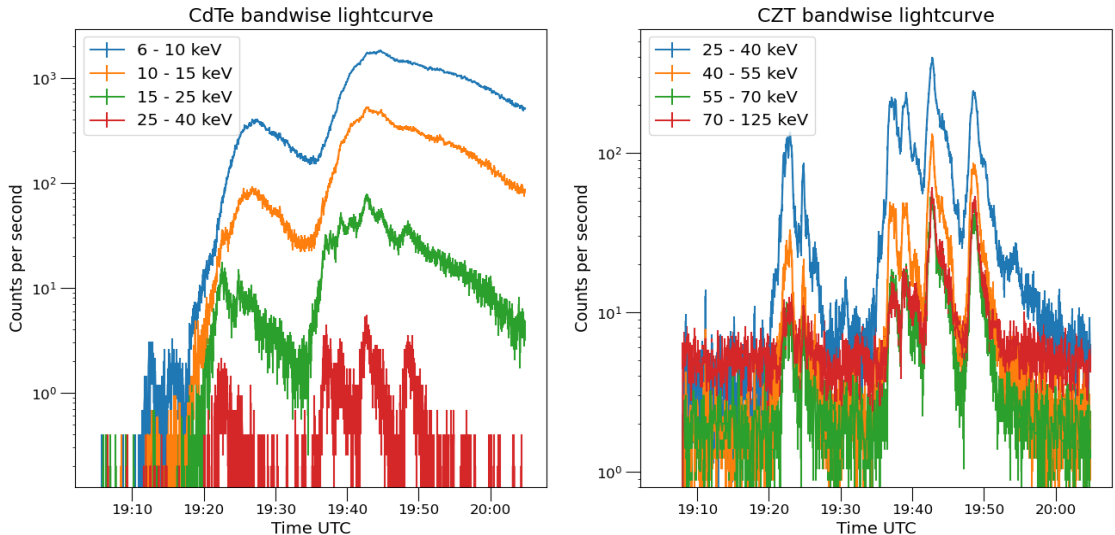

**Figure 13**: Energy-dependent X-ray light curves from CdTe and CZT detectors during a flaring event (X-class) on 28 November 2023. Oscillatory signatures in the flare are clearly seen in high-energy band light curves above 15 keV.

The time-resolution for each event recorded by the instrument is 10 ms. This allows for high-cadence light curves during highly intense hard X-ray flares. In Figure 13, we show energy-dependent light curves detected by the CdTe and CZT detectors. This light curve corresponds to the X-class flare (as per GOES classification) detected on 28 November 2023. This is the first event after switching on that has a significant count rate in both types of detectors, covering the entire energy range. The QPP feature is observed in higher energy bands above 15 keV. Oscillatory features are not obvious in the CdTe spectra (below 15 keV), whereas it is very prominent in higher energy bands (above 25 keV) of CZT spectra.

In Figure 14, we show the combined spectrum (7 – 150 keV) of X-class flare during the peak of the flare with an integration time of 60 sec. The spectrum clearly demonstrates the capabilities of both detectors of spectral range of 7 to 150 keV. The spectral modeling of the combined spectra will be presented elsewhere.

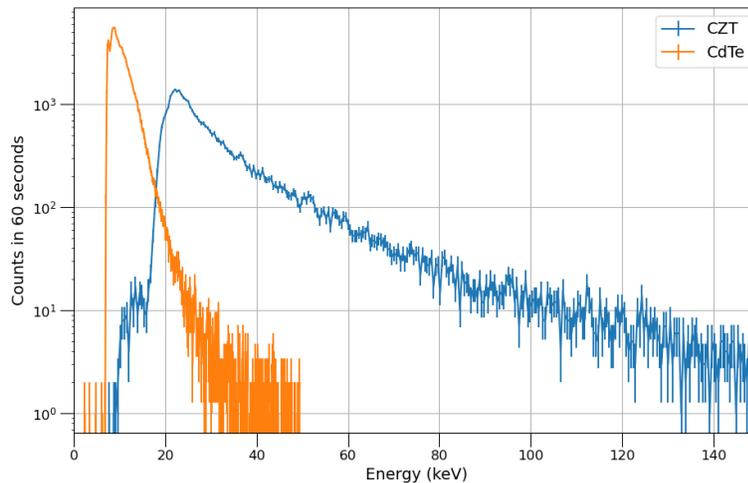

**Figure 14**: Combined X-ray spectrum (7 keV – 150 keV) from CdTe and CZT detectors for a 60-second integration around the X-class flare peak at 19:42 UT on 28 November 2023.

By default, the observed data is downlinked once a day from the spacecraft. At the Indian Space Science Data Centre (ISSDC), the raw data downloaded at the ground station are converted to Level-0, packing the payload data with auxiliary data like the time correlation table and the Spacecraft, Planet, Instrument, C-matrix and Events (SPICE) kernels. The Level-0 data made available at the data server of ISSDC are transferred to the Payload Operation Centre (POC) at the ISITE campus of URSC, where the data pipeline processes these to generate higher-level products – Level-1 and Level-2 data. The primary Level-1 data products of HEL1OS consist of FITS files compliant with the Office of Guest Investigators Program (OGIP) of the National Aeronautics and Space Administration (NASA): (a) type II PHA spectral files for each of the detectors, at 20 sec cadence, and (b) the light curves for each detector at 1-sec cadence in different energy sub-bands saved in different table extensions. Other files made available as part of the Level-1 data in FITS format are: the event-list file (in which all the detected events, recorded at the instrument clock resolution of 10 ms in each of the detectors, are saved in different table extensions), and other auxiliary files, e.g., data related to the good time intervals, housekeeping parameters – all in FITS format. The format of Level-2 primary data products, i.e., light curves and spectra, is the same as for Level-1 products, but are generated after correcting for the pile-up in the CdTe events, and the CZT events are filtered for saturation. These are packed along with the relevant response files for spectral analysis and disseminated on the portal of ISSDC called **P**olicy based data **R**etrieval, **A**nalytics, **D**issemination **A**nd **N**otification system (PRADAN) for Aditya-L1 mission. Also provided on PRADAN portal are some Python-based utility tools that will help in the timing and spectral analysis of the HEL1OS data. More details of the data products are planned in a separate publication.

## 7. Summary and Conclusion

HEL1OS on board Aditya-L1 is a dedicated hard X-ray spectrometer to study the solar flare from the Sun. The science goals of HEL1OS will benefit greatly from complementary measurements by other instruments on Aditya-L1 like the VELC, SoLEXS, and SUIT, as well as with ground-based observations. The instrument is designed with two different types of X-ray detectors namely CdTe and CZT. This also includes an in-house development of low-noise electronics which is one of the key developments towards the realization of the instrument. Many of the developed technologies for HEL1OS will be useful for the future development of X-ray instruments. HEL1OS has been operating successfully since January 2024 after the insertion of the Aditya-L1 spacecraft into Halo orbit.

Since the commissioning of the instrument, HEL1OS has detected all the activity that is reported by GOES bands, and sometimes beyond them as HEL1OS detectors operate in harder X-ray bands. To date, the flares detected by HEL1OS with statistically significant spectra can be associated with GOES classes C6 – X3. Preliminary (in-flight) results along with achieved specifications are in line with expected performance. From 1 January 2024, till the end of June 2025, there have been a large number of flares of GOES classes C3 and above up to X-class (Ravishankar et al., 2025, in preparation). The operating temperatures and other parameters of the payload have been well-constrained within the designed ranges. In addition, owing to the payload mounting within the spacecraft deck, and also the spacecraft itself orbiting about the L1 Lagrangian point away from the Earth's radiation belts, the background level of the data has been very benign.

## Acknowledgement

The authors thank the anonymous reviewer for constructive comments and useful suggestions that helped to improve the quality of the manuscript. HEL1OS is designed and developed at the Space Astronomy Group (SAG) of U R Rao Satellite Centre (URSC), Indian Space Research Organisation along with the various entities within URSC. The Thermal Systems Group provided the required thermal design, analysis and implementation support. The Systems Integration Group helped in the structural analysis of the payload and participated in the vibration tests. The Reliability and Quality Assurance Group ensured that all non-space grade components are qualified or up-screened for its reliability and quality. Space environmental tests are conducted at various laboratories of the facilities within URSC. The Assembly and Integration team took the responsibility of integrating the payload to the satellite. The Mission team helped in the L0 data generation including the required HK and SPICE data of Aditya-L1. The ISRO Satellite TRAcking Centre (ISTRAC) is the nodal point for all operations and the Indian Space Science Data Centre (ISSDC) within ISTRAC handles data downloading from the spacecraft and communication with the Payload Operation Centre (POC). They will be the dissemination centre for the science ready products. The Aditya-L1 project team coordinated the overall activities as a nodal point. The spectral calibration of the CdTe detectors was undertaken at the INDUS-2 beamline facility of the Raja Ramanna Center for Advanced Technology (RRCAT) at Indore. The POC of HEL1OS is at SAG of URSC, where the higher-level data product is generated and posted to ISSDC for dissemination to the scientific community. Authors (SAG) thank the Group Head, SAG, Deputy Director, Payload Data Mgmt. and Space Astronomy Area, Associate Director and Director of URSC for encouragement and continuous support to carry out this work.

HEL1OS on Aditya-L1Aditya-L1 is an observatory class mission that is fully funded and operated by the Indian Space Research Organisation. The mission was conceived and developed with the help of all ISRO Centres, and the payloads were done by the payload PI Institutes in close collaboration with ISRO and many other national institutes such as the Indian Institute of Astrophysics (IIA), Inter-University Centre of Astronomy and Astrophysics (IUCAA), Laboratory for Electro-Optics System (LEOS) of ISRO, Physical Research Laboratory (PRL), U R Rao Satellite Centre of ISRO and Vikram Sarabhai Space Centre (VSSC) of ISRO.

Hester, J. J.: 2008, The Crab Nebula: an astrophysical chimera. *Annual Rev. Astron. Astrophys.* **46**, 127

Holman et al.: 2011, Implications of X-ray Observations for Electron Acceleration and Propagation in Solar Flares, *Space Sci. Rev.* **159**, 107

Hudson, H. S. et al.: 2021, Hot X-ray onsets of solar flares. *Mon. Not. R. Astron. Soc.* **501**, 1273

Inglis, A. R. et al.: 2016, A Large-Scale Search for Evidence of Quasi-Periodic Pulsations in Solar Flares. *Astrophys. J.* **833**, 284

Jain, A. et al.: 2025, High Energy L1 Orbiting X-ray Spectrometer (HEL1OS) onboard Aditya-L1: Challenges and Novel Developments. *J. Space Tech.*, Under review

Jelínek, P. & Karlický, M.: 2019, Pulse-beam heating of deep atmospheric layers, their oscillations and shocks modulating the flare reconnection. *Astron. Astrophys.* **625**, A3

Joshi, B. et al.: 2011, Pre-flare Activity and Magnetic Reconnection during the Evolutionary Stages of Energy Release in a Solar Eruptive Flare. *Astrophys. J.* **743**, 195

Joshi, B. et al.: 2016, Pre-flare coronal jet and evolutionary phases of a solar eruptive prominence associated with the M1.8 flare: SDO and RHESSI observations. *Astrophys. J.* **832**, 130

Knoll, G. F.: 2000, Radiation Detection and Measurement. *John Wiley & Sons, Inc.*, 3$^{rd}$ Edition

Knuth, T. & Glesener, L. et al.: 2020, Subsecond Spikes in Fermi GBM X-ray Flux as a Probe for Solar Flare Particle Acceleration. *Astrophys. J.* **903**, 63

Kontar, E.P., Dickson, E. & Kašparová, J.: 2008, Low-Energy Cutoffs in Electron Spectra of Solar Flares: Statistical Survey. *Sol. Phys.* **252**, 139

Kontar et al.: 2011, Deducing Electron Properties from Hard X-ray Observations, *Space Sci. Rev.* **159**, 301.

Kotoch, T. et al.: 2011, Instruments of RT-2 experiment onboard CORONAS-PHOTON and their test and evaluation II: RT-2/CZT payload. *Exp. Astron.* **29**, 27

Krucker, S. et al.: 2020, The Spectrometer/Telescope for Imaging X-rays (STIX). *Astron. Astrophys.* **642**, A15

Kushwaha, U. et al.: 2015, Large-scale contraction and subsequent disruption of coronal loops during various phases of the M6.2 flare associated with the confined flux rope eruption. *Astrophys. J.* **807**, 101

Li et al.: 2024, Localizing quasi-periodic pulsations in hard X-ray, microwave and Lyα emissions of an X6.4 Flare. *Astrophys. J.* **970**, 77

Li, D. et al: 2024, A statistical investigation of the Neupert effect in solar flares observed with ASO-S/HXI, *Sol. Phys.* **299**, 57.

Lin, R. P. et al.: 2002, The Reuven Ramaty High-Energy Solar Spectroscopic Imager (RHESSI). *Sol. Phys.* **210**, 3

Nandi et al.